# Fractal Structure of Shortest Paths in Native Proteins and Determination of Residues on a Given Shortest Path

Burak Erman

*Department of Chemical and Biological Engineering, Koc University, Istanbul, Turkey*

email: berman@ku.edu.tr

Fractal structure of shortest paths depends strongly on interresidue interaction cutoff distance. The dimensionality of shortest paths is calculated as a function of interaction cutoff distance. Shortest paths are self similar with a fractal dimension of 1.12±0.03 when calculated with step lengths larger than 6.8 Å. Paths are multifractal below 6.8 Å. The number of steps to traverse a shortest path is a discontinuous function of cutoff size at short cutoff values, showing abrupt decreases to smaller values as cutoff distance increases. As information progresses along the direction of a shortest path a large set of residues are affected because they are interacting neighbors to the residues of the shortest path. Thus, several residues are involved diffusively in information transport which may be identified with the present model. An algorithm is introduced to determine the residues of a given shortest path. The shortest path residues are the highly visited residues during information transport. These paths are shown to lie on the high entropy landscape of the protein where entropy is taken to increase with abundance of visits to nodes during signal transport.

# Introduction

Flow of energy or information in general, from one point to another in native proteins depends on the spatial arrangement of amino acid residues. There is a close relationship between fractal structure and information transport in proteins which has been studied by several authors and reviewed by Leitner [1]. For a dense, homogeneous system, energy may flow in three dimensions which is characterized by the spectral dimension of 3 obtained from the celebrated Debye density of states expression. For such media, the mass dimension which is the Euler dimension is also 3. The three dimensional structures of proteins are neither dense nor homogeneous, and consequently the dimensionality for the propagation of energy, i.e., the spectral dimension, and the mass fractal dimension both deviate from 3. The value is around 1.7 for the spectral dimension [2-4] and 2.5 for mass fractal dimension [5]. The loss in spectral dimension results from the fact that not every path is allowed for a signal to proceed from one point to the other in the protein. Similarly, the loss in mass fractal dimension is because one can only travel in the protein from atom to atom, and the problem is similar to random walk on lattices[1, 6]. However, in contrast to random walk, information flow is anisotropic in proteins and the possible routes in the protein are constrained by the presence of irregularities in mass density. Moreover, the specific structure of a protein evolved for fulfilling and optimizing a specific function is expected to impose constraints on flow of energy, such as going over the shortest path, shortest time, least resistance, through highly visited nodes, or combinations of these. In allosteric information transport, for example, it is plausible to expect flow along the shortest path in the fractal architecture for which the transit time, quality, robustness, and noise-independence of signal transmission should be of major concern. Imposing motion to follow shortest paths rather than diffuse over space reduces the dimensionality significantly to values between the upper limit of 3 and the lower limit of 1, The latter is the dimension of the shortest path in homogeneous Euclidean space. In this paper, based on residue contact maps of proteins, we analyze the physics of shortest paths, their fractal dimensions, and present a computational algorithm for determining the residues on a given shortest path.

Among several possible paths between two points i and j in a protein, there exists one or a small number of paths that carry the information from i to j in the smallest number of steps for a given step size. The step size depends basically on the range of interactions in the structure, the fractal structure that determines the interacting neighbors, and the characteristic length scale of the perturbation. We emphasize the dependence of interactions on step size throughout the paper. In the following section, we give a precise definition of shortest path between two points in proteins and outline the calculation of shortest path dimension and the Hausdorff dimension. The identification of residues that lie on a given path is an important problem since it is usually the desire to understand their behavior and manipulate them. We outline a method for determining the residues that lie on a given shortest path. We also show that the path residues are the highly visited residues among the larger set of residues that are indirectly perturbed in the neighborhood as signal propagates. This has a thermodynamic significance indicating that the shortest path among others is a low entropy barrier path.

The characteristic packing of helix-beta-coil structures exhibits an unusual dependence on step size. The number of steps to span a path is obviously very large for very small step sizes. The decrease in the number of steps with increasing step size is not continuous, however. An infinitesimal increase in step size may result in a sudden drop in the number of steps. This behavior is representative of an abrupt change of regime which may have significance when small wavelength perturbations propagate in the fractal structure of the protein. For step sizes in the order of the first coordination shell radius and larger, shortest paths have dimensionality of around 1.1, surprisingly close to that of Euclidean paths. Our calculations are based on a large number of paths from several proteins. All calculated entities show the same features as outlined in more detail below.

# Methods and Materials

Unlike dense homogeneous continuous media, a protein is discontinuous, locally anisotropic an inhomogeneous [7] in which signal travels on a contact map in steps. A signal can go from a point to its neighboring point in one step if the two points are interacting directly, i.e., if they are 'connected' in the language of graph theory. The connection is defined in terms of an interaction radius whose magnitude is kept as a variable in this analysis. Its choice directly affects the entries of the contact matrix and the physics based on the contact matrix. The interactions may depend on the characteristic wavelength of the signal. High frequency perturbations may activate the interactions of nearby atoms whereas low frequency ones may activate distant neighbors.

The shortest interaction path between two points i and j is obtained when the signal goes from i to j in a minimum of n steps of a given step size. At lengthscales of the order of first coordination shell and below, which are the relevant scales for signal transmission, native proteins are not self-similar objects (See below). A multitude of fractal dimensions are active at these scales and their detailed understanding is needed for predicting and controlling protein function, most importantly for determining allosteric pathways and modifying them. For long wavelength perturbations, a single fractal dimension is obtained for these paths, which is determined basically by the contact topology and is otherwise independent of the detailed three dimensional structures of the proteins. Fractal structure of native proteins has been addressed in several papers [1, 5, 6, 8-14]. Here, we focus on fractal properties of shortest interaction paths in proteins, and in this respect it is a new addition to the existing knowledge on protein structure and function. We keep the term 'interaction' general, which may result from any source of perturbing effect.

We let $r_c$ denote the interaction cutoff radius and treat it as variable in order to understand the dependence of fractal behavior on different lengthscales of interaction. A perturbation with a wavelength less than $r_c$ generates interactions within a sphere of radius $r_c$. The ij'th element of the contact matrix $C$ is 1 if the points i and j are within $r_c$ and zero otherwise. We adopt the alpha carbon representation of the protein. Two alpha carbons do not interact directly if they are at a distance larger than $r_c$. Below, we obtain the fractal dimension of shortest interaction paths in proteins as a function of $r_c$.

*Fractal Dimension of Shortest Interaction paths*: The emphasis is on finding the shortest path among a multitude of paths between two points on the protein. Starting from a residue i, one can go to residue j following a path of interacting residues. We are interested in finding a path traversed with a minimum number of steps, each step being defined as hopping from one point to its interacting neighbor. One can go from i to j in $k$ steps in $m$ different ways. In terms of the contact matrix, $m = \left[C^k\right]_{ij}$, i.e., the ij'th element of the k'th power of $C$. The minimum number of steps to go from i to j, $k_{min}$, is obtained when $\left[C^{(k-1)}\right]_{ij} = 0$ and $\left[C^k\right]_{ij} \neq 0$ [15]. The minimum number of steps $k_{min}$ will be a function of $r_c$. In an Euclidean dense homogeneous uniform object, the path will be a straight line from i to j, and we will have $r_{ij} = k_{ij,\min} r_C$, where $r_{ij}$ is the Euclidean distance between i and j. For the protein, this will not hold due to fractal morphology. The minimum number of steps in going from i to j will be expressed in terms of a power law: $r_{ij}^D = A k_{ij,\min} r_C$, where $D$ is the fractal dimension and $A$ is the constant of proportionality of dimension $D-1$. For long paths, we show in the Supplementary Material Section 1 that $D$ is written as

$$D = \frac{\log(k_{ij,\min} r_C)}{\log(r_{ij})} \tag{1}$$

This definition of dimension will strictly hold for a large system and will read as $D = \lim_{\frac{r_{ij}}{r_C} \to \infty} \frac{\log(k_{ij,\min} r_C)}{\log(r_{ij})}$. For proteins with finite dimensions, Eq. 1 serves as a good approximation to dimension as shown below.

The Hausdorff dimension $D_H$ of fractal paths is obtained from the expression [16]

$$D_H = \lim_{k \to \infty} \frac{\log(N_k)}{\log(1/r_k)} \tag{2}$$

where the path is partitioned into $N_k$ segments of length $r_k$ each. For the shortest path, we identify the length of segments $r_k$ with the cutoff distance $r_C$ and $N_k$ with $k_{ij,\min}$.

*Determining the residues along the shortest path:*

Given the end residues, $i_{\{1\}}(0)$ and $i_{\{1\}}(N)$. The numbers in parenthesis give the step number along the directed walk from the starting to the ending residue and the subscript indices in braces give the set of residues that are neighbors to the residues involved in the previous step. Thus, the set $\{j\}$ at an intermediate point will contain more than one residue. The residues involved at each step are candidate residues that contain the residue of the shortest path which will be identified during backtracking (See below). In the first step, starting from $i_{\{1\}}(0)$ we find the set of residues $i_{\{j\}}(1)$ that can be accessed from $i_{\{1\}}(0)$. The residues $i_{\{j\}}(k)$ are accessed from the residues of the previous step $i_{\{j\}}(k-1)$. Residue j at the kth step will be accessed by $v_{\{j\}}(k)$ different paths from residues of the previous step. The set of resides involved in going from $i_{\{1\}}(0)$ to $i_{\{1\}}(N)$ is $\{i_{\{1\}}(0), i_{\{j\}}(1), i_{\{j\}}(2), \ldots, i_{\{j\}}(k-1), i_{\{j\}}(N)\}$ and the number of arrivals to the residues from previous paths during the forward trip is $\{1_{\{j\}}(1), v_{\{j\}}(2), \ldots, v_{\{j\}}(k-1), v_{\{j\}}(N)\}$. The residues of the shortest path, $\{i_0, i_1, i_2, \ldots, i_{k-1}, i_k, \ldots, i_N\}$ are obtained by backtracking, starting from N and going step by step to 1. When the residue at step k is identified in this way, the residue at step k-1 is determined from the list of residues that are adjacent to residue k, already determined in the previous step. This process of backtracking may lead to more than one shortest path, all of which may be identified individually by the backtracking method. Sample calculations for the path from LYS132 to ASP98 in 1GPW.pdb are presented in Supplementary Material Section 2.

*Entropy in signal transport:*

Identification of the shortest path residues also leads to the set $\{v_1, v_2, \ldots, v_{k-1}, v_k, \ldots, v_N\}$ identifying the number of visits to each residue along forward tracking, i.e., along the traveling direction of information. The frequency of visits to the i'th residue is written as $f_i = \dfrac{v_i}{\sum v_i}$ where the sum in the denominator for all visits to all residues involved in forward tracking. The denominator here is representative of the space available to the forward tracking process. The entropy of information transport over the shortest path may then be defined as $S/k = -\sum f_i \log f_i$ where the sum is over the residues of the shortest path. According to this definition, the total entropy of transport using all paths is the sum of entropies of transport through individual paths. It is plausible that a signal will choose high entropy paths in the absence of imposed additional constraints. It should be added that since the shortest path is directional, i.e., the nodes visited in going along the forward and reverse directions are different, the entropies of signal transport in the forward and reverse directions will be different.

## Results

Values of $D$ from Eq. 1 are calculated for 150 paths from ten proteins from the Protein Data Bank ( 3K8Y, 1SHJ, 4FHB, 2V20, 2V1Z, 3PGF, 4EZF, 4H7U, 1GPW, 4N4J) and presented in Figure 1 as a function of $r_C$. These proteins are known to participate in allosteric activity. For each of the 10 proteins 15 paths are chosen such that $r_{ij} \geq 40$ Å for each path, summing up to 150 paths in total. The filled circles represent the average of the 150 paths. Unfilled circles represent the averages for each of the 15 paths from each protein.

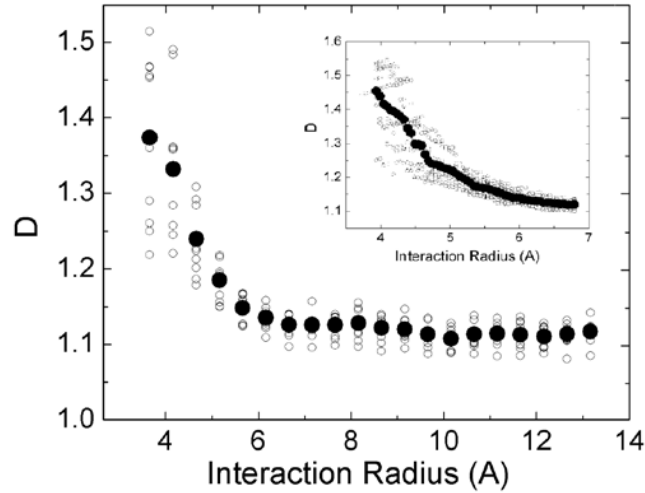

FIG 1. Fractal dimensions of interaction paths in proteins as a function of interaction cutoff radius. Filled circles show averages for 150 different paths from 10 proteins. Empty circles show averages for 15 paths from each protein. The lowest bond percolation limit is ca. 3.8 Å below which a path does not exist in the alpha carbon representation. Inset shows details at smaller interaction radii.

Generally, fractal objects are characterized by a single number [16] which is true for self similar objects. Proteins are not self similar at every length scale as can be seen from the figure. Above lengthscales of $r_C \approx 6.8$ Å, interaction paths have single fractal dimension of $D = 1.12 \pm 0.03$ obtained from Figure 1. However, below $r_C \approx 6.8$ Å, the protein is multifractal; exhibiting different values of $D$ for different interaction radii, starting from average values of $D = 1.4$ at the percolation threshold of ca. 3.8 Å. A closer look to the multifractal range, the inset in Figure 1, surprisingly shows that there is a fractal transition at a length scale slightly below 4.5 Å. At lengthscales below this limit, the system traverses the path between points i and j in a large number of steps, but at the transition point, this number exhibits a jump to lower values. This behavior is observed for all of the ten proteins studied. The magnitude of the drop and the corresponding interaction radius differ from protein to protein, however. This transition is due to the sharp decrease in the number of steps in going from the initial to the final point on the path as the step length increases. This is shown in Figure 2 for the protein 1GPW.pdb for the known allosteric path[17, 18] from LYS132 to ASP98 (See Supplementary Material for more details). LYS 132 is at the allosteric site to which a ligand binds and the effect of binding is transmitted to ASP98 which is 29 Å away. In fact there is more than one sharp drop, closely spaced in the range 4 to 5 Å. This is also true for all the other proteins studied.

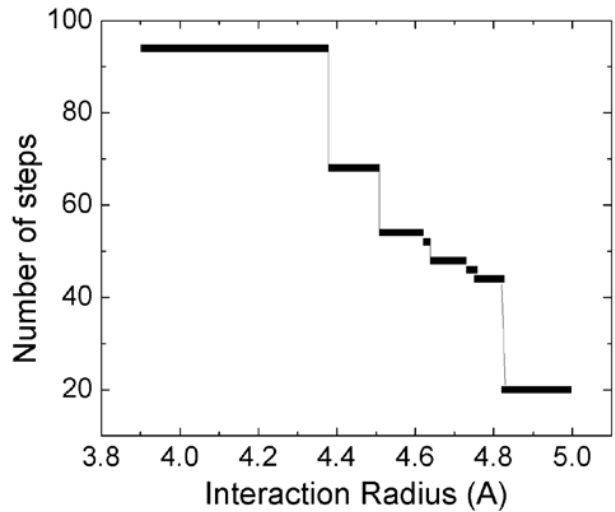

FIG 2. Discontinuities of number of steps at small wavelengths for 1GPW.pdb for the shortest path from residue LYS132 to ASP98.

In Figure 3, we present results of calculations of the Hausdorff dimension calculated using Eq. 2 for 150 paths from the set of ten proteins. The straight line is the least squares fit to the data points. Its slope, which approximates the Hausdorff dimension given by Eq. 2 is $D_H = 1.1$, which is approximately the fractal dimension obtained in the preceding section. The maximum abscissa value of -1.9 corresponds to a cutoff radius of 6.8 Å and the minimum ordinate value of -2.4 corresponds to 11 Å. The points outside this range are not presented in the figure because below 6.8 Å, the jumps shown in Figure 2 lead to discontinuous data, and above 11 Å the system reaches a saturation level [16] and the slope equates to unity.

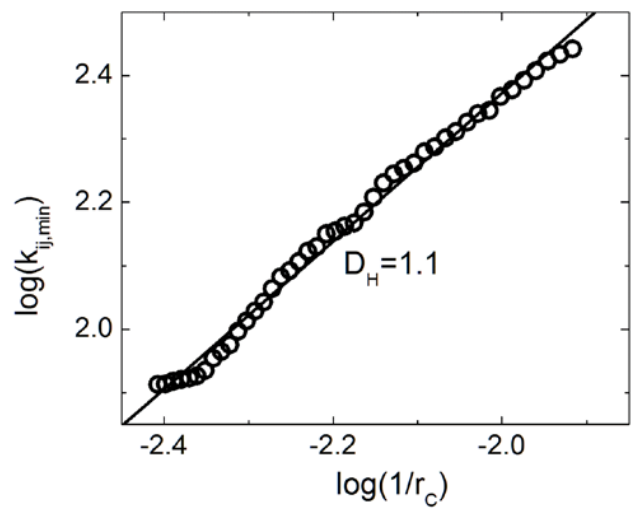

FIG 3. Hausdorff dimension obtained as the average from 150 paths in 10 proteins.

In Figure 4, we present the residues visited during signal propagation for the allosteric shortest path for 1GPW.pdb from residue LYS132 to ASP98. The shortest path with a cutoff radius of 6.9 Å reaches ASP98 in six steps (See Supplementary Material). Abscissa values indicate residues indices and ordinate values show the frequency of visits to the residues while going from the initial point to the final on the path, i.e., in going from LYS132 to ASP98. As the different ordinate values indicate, some residues will be visited more frequently than others. Several paths from LYS132 to ASP98 are possible, but only one set of residues gives the shortest path, shown by the dark circles identified during backtracking. The importance of these residues in allosteric communication for 1GPW.pdb has been shown before [17,18]. The path residues are situated in the highly visited regions in the figure. The list of residues visited during forward tracking is given in the Supplementary Material Part 2 and the residues on the shortest path, obtained during backtracking are highlighted in that list.

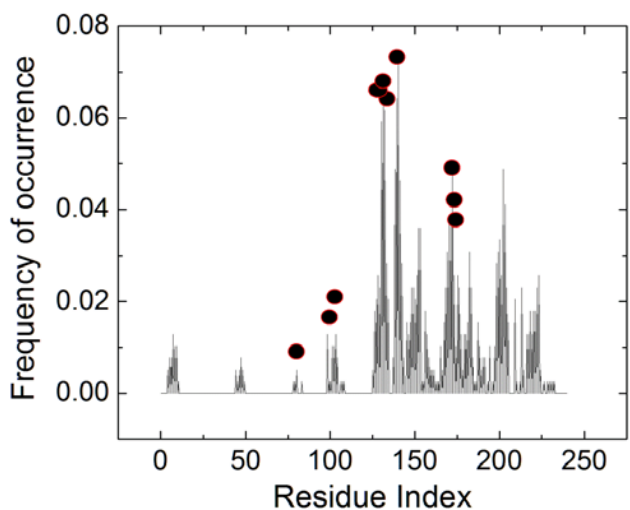

FIG 4. Residues on all paths from LYS132 to ASP98 for 1GPW.pdb. Shortest path residues are indicated by filled circles.

Calculations over $10^5$ random paths of six steps, starting from LYS132 and terminating at ASP98 showed that the entropy of the shortest path is larger than 95% of the entropies of the random paths. Thus, in the absence of constraints, signal transport prefers high entropy routes through the protein.

# Discussion and Conclusions

We studied fractal dimensions of shortest paths as a function of the interaction cutoff radius. The transition of the fractal dimension from multifractal to self similar regimes takes place at 6.8 Å which equates to the radius of the first coordination shell in the alpha carbon representation of native proteins. The number of steps to go between two points exhibits sharp discontinuities around cutoff lengths of 4 Å. Below 3.8 Å the system is non-percolating, as in the bond percolation problem, and a path between two points does not exist. The fractal dimension of shortest paths at and above the cutoff value of 6.8 Å is $D=1.12\pm0.03$, remarkably close to $D=1$ of Euclidean dense homogeneous systems where the shortest distance between two points is a straight line. The Hausdorff dimension of shortest paths is 1.1, close to the value obtained by using Eq. 1. In step sizes much less than 6.8 Å information travels in a diffusive manner with higher dimensionalities of paths carrying this information. If the signal were not directional, then it would be equal to 2.5 from the work of Enright and Leitner at the shortest wavelength [5] instead of 1.4 from Figure 1. Perturbing a residue perturbs, in turn, several residues directly connected to it by the network structure of the protein. Thus the perturbations spread into the protein and a multitude of perturbation pathways of differing lengths, originating from a point and ending at another, becomes possible. We proposed an algorithm to find the residues along the shortest of these paths. Knowledge of the residues along a path may have important ramifications, particularly in understanding and controlling allosteric communication or even non-allosteric signal transport[19]. One may make the plausible assumption that allosteric information travels on a shortest path from the effector site to the active site. A detailed study of this conjecture requires optimization of the variables imposed by evolution. An excellent example of constrained optimization to calculate fractal dimensions is by West et al, who calculated the scaling exponents for biological systems that optimize the surface area to volume ratios [20]. For shortest path scaling, we do not know yet what the constraints for information flow are. However, the present calculations show that shortest path residues lie mostly on high entropy routes. The present calculations also show that entropy of information transport from the effector to active site is not equal to that in the reverse direction, leading to the conclusion that allosteric communication is directional.

The present analysis is based on information from the contact map of proteins with a preselected interaction cutoff radius. The emergence of the cutoff radius approximately as 6.8Å above which the paths are mono-fractal is not a coincidence. That this value of the cutoff is the value for which the Gaussian Network Model (GNM) [21] performs best has been shown in a large number of papers. The spectral dimension for slow vibrational modes from GNM with an $r_c$ of 7.0Å, was shown to be 1.63 in contrast to 3 of the Debye theory [3]. The relationship between spectral dimension and mass dimension of three dimensional objects, including proteins, has been widely studied. In the present work, we were able to establish an equality of the short path dimensions with Hausdorff dimensions only. Unlike the GNM work, the fractal dimension of 1.1 is structural only and is not a spectral quantity. Spectral dimension of shortest paths, which will be less than or equal to 1.1, should be calculated with a dynamic model of transmission over a path and awaits future work.

The present calculations are based on the alpha carbon coarse grained approximation which should suffice for a proof of principle approach. Although atomic detail should not change the overall picture, specific values may slightly depend on the coarseness of atomic organization as has been shown in the calculations of mass fractal dimensions of proteins [5].

As a final remark, the present study of signal flow focuses on the structural view of allosteric communication only and does not consider the enthalpic side of the phenomenon. Although a large number of papers show the importance of structure on allosteric transmission[18, 22-26] the structural versus enthalpic picture has recently been brought into attention by Motlagh et. al. [27], which definitely deserves attention. In this respect, the present model is only a partial picture of the complete process of signal transport in proteins.

Acknowledgment: The author gratefully acknowledges fruitful discussions with Professor Ahmet Gul, School of Medicine, Istanbul University.

**SUPPLEMENTARY MATERIAL**

# 1. Derivation of Eq. 1

Taking the logarithm of both sides of the expression $r_{ij}^D = A k_{ij,\min} r_C$ we obtain

$$\log(k_{ij,\min} r_C) = -\log(A) + D \log(r_{ij}) \tag{S1}$$

Taking $\log(k_{ij,\min} r_C)$ as the y-axis and $\log(r_{ij})$ as the x-axis, Eq 1 plots as a straight line with $-\log(A)$ as the y-intercept and D as the slope. The slope is written for points 1 and 2 as

$$D = \frac{\log(k_{ij,\min}(2) r_C) - \log(k_{ij,\min}(1) r_C)}{\log(r_{ij}(2)) - \log(r_{ij}(1))} \tag{S2}$$

Now we choose a path 1 traversed with $k_{ij,\min}(1)$ steps and another path that is traversed with $k_{ij,\min}(1) + 1$. The length of the first and second paths will be $r_{ij}(1)$ and $r_{ij}(1) + \delta r_{ij}$, respectively, with $r_{ij}(1) \gg 1$ and $r_{ij}(1) \gg \delta r_{ij}$ for large $r_{ij}$ we can write

$$\frac{\log(k_{ij,\min}(2) r_C)}{\log(r_{ij}(2))} = \frac{\log((k_{ij,\min}(1)+1) r_C)}{\log(r_{ij}(1) + \delta r_{ij})} \approx \frac{\log(k_{ij,\min}(1) r_C)}{\log(r_{ij}(1))} = \frac{\log(k_{ij,\min}(2) r_C) - \log(k_{ij,\min}(1) r_C)}{\log(r_{ij}(2)) - \log(r_{ij}(1))} \equiv D \tag{S3}$$

# 2. Sample Calculations for 1GPW.pdb

List of residues visited during the six steps in going from LYS132 to ASP98 in 1GPW.pdb. At the kth step, the left column in the following list is the residue in step k-1, the second column is the residue in the kth step. The third column shows the number of different ways a path arrives to the shown residue at the kth step. The highlighted residues are the ones obtained during backtracking. For this, we start from residue 98 at step 6 and identify its path neighbor at step 5. Then, proceed to finding the path neighbor at step 4, etc. Backtracking in this way may result in more than one path which is listed at the bottom of this list. Finally, the residues on the shortest path 1 are shown on the protein.

| Residue at step k-1 | Residue at step k | Number of paths |
|---|---|---|
| **Step 1** | | |
| 132 | 138 | 1.000000 |
| 132 | 139 | 1.000000 |
| 132 | 140 | 1.000000 |
| 132 | 172 | 1.000000 |
| 132 | 173 | 1.000000 |
| 132 | 174 | 1.000000 |
| **Step 2** | | |
| 138 | 131 | 5.000000 |
| 138 | 132 | 6.000000 |
| 138 | 133 | 2.000000 |
| 138 | 134 | 2.000000 |
| 138 | 152 | 3.000000 |
| 138 | 153 | 3.000000 |
| 139 | 131 | 5.000000 |
| 139 | 132 | 6.000000 |
| 139 | 133 | 2.000000 |
| 139 | 134 | 2.000000 |
| 139 | 150 | 2.000000 |
| 139 | 151 | 2.000000 |
| 139 | 152 | 3.000000 |
| 139 | 153 | 3.000000 |
| **140** | **130** | **3.000000** |
| 140 | 131 | 5.000000 |
| 140 | 132 | 6.000000 |
| 140 | 148 | 1.000000 |
| 140 | 149 | 1.000000 |
| 140 | 150 | 2.000000 |
| 140 | 151 | 2.000000 |
| 140 | 152 | 3.000000 |
| 140 | 153 | 3.000000 |
| **172** | **130** | **3.000000** |
| 172 | 131 | 5.000000 |
| 172 | 132 | 6.000000 |
| 172 | 176 | 3.000000 |
| 172 | 183 | 1.000000 |
| 172 | 202 | 1.000000 |
| **173** | **130** | **3.000000** |
| 173 | 131 | 5.000000 |
| 173 | 132 | 6.000000 |
| 174 | 132 | 6.000000 |

Step 3

| | | |
|---|---|---|
| 130 | 140 | 30.000000 |
| 130 | 141 | 13.000000 |
| 130 | 145 | 5.000000 |
| 130 | 170 | 7.000000 |
| 130 | 171 | 14.000000 |
| 130 | 172 | 29.000000 |
| 130 | 173 | 26.000000 |
| 131 | 138 | 25.000000 |
| 131 | 139 | 31.000000 |
| 131 | 140 | 30.000000 |
| 131 | 141 | 13.000000 |
| 131 | 171 | 14.000000 |
| 131 | 172 | 29.000000 |
| 131 | 173 | 26.000000 |
| 132 | 138 | 25.000000 |
| 132 | 139 | 31.000000 |
| 132 | 140 | 30.000000 |
| 132 | 172 | 29.000000 |
| 132 | 173 | 26.000000 |
| 132 | 174 | 16.000000 |
| 133 | 137 | 7.000000 |
| 133 | 138 | 25.000000 |
| 133 | 139 | 31.000000 |
| 134 | 138 | 25.000000 |
| 134 | 139 | 31.000000 |
| 137 | 133 | 13.000000 |
| 138 | 131 | 21.000000 |
| 138 | 132 | 19.000000 |
| 138 | 133 | 13.000000 |
| 138 | 134 | 13.000000 |
| 138 | 152 | 11.000000 |
| 138 | 153 | 9.000000 |
| 139 | 131 | 21.000000 |
| 139 | 132 | 19.000000 |
| 139 | 133 | 13.000000 |
| 139 | 134 | 13.000000 |
| 139 | 150 | 10.000000 |
| 139 | 151 | 12.000000 |
| 139 | 152 | 11.000000 |
| 139 | 153 | 9.000000 |
| 140 | 130 | 15.000000 |
| 140 | 131 | 21.000000 |
| 140 | 132 | 19.000000 |
| 140 | 148 | 6.000000 |
| 140 | 149 | 8.000000 |
| 140 | 150 | 10.000000 |
| 140 | 151 | 12.000000 |
| 140 | 152 | 11.000000 |
| 140 | 153 | 9.000000 |
| 141 | 130 | 15.000000 |
| 141 | 131 | 21.000000 |
| 141 | 145 | 5.000000 |
| 141 | 146 | 2.000000 |
| 141 | 147 | 4.000000 |
| 141 | 148 | 6.000000 |
| 141 | 149 | 8.000000 |
| 148 | 140 | 30.000000 |
| 148 | 141 | 13.000000 |
| 148 | 142 | 4.000000 |
| 149 | 140 | 30.000000 |
| 149 | 141 | 13.000000 |
| 149 | 142 | 4.000000 |
| 150 | 139 | 31.000000 |
| 150 | 140 | 30.000000 |
| 151 | 139 | 31.000000 |
| 151 | 140 | 30.000000 |
| 151 | 156 | 8.000000 |
| 152 | 138 | 25.000000 |
| 152 | 139 | 31.000000 |
| 152 | 140 | 30.000000 |
| 152 | 156 | 8.000000 |
| 153 | 138 | 25.000000 |
| 153 | 139 | 31.000000 |
| 153 | 140 | 30.000000 |
| 153 | 157 | 3.000000 |
| 130 | 128 | 4.000000 |
| 170 | 129 | 11.000000 |
| 170 | 130 | 15.000000 |
| 170 | 198 | 1.000000 |
| 170 | 199 | 1.000000 |
| 170 | 200 | 3.000000 |
| 170 | 201 | 4.000000 |
| 171 | 129 | 11.000000 |
| 171 | 130 | 15.000000 |
| 171 | 131 | 21.000000 |
| 171 | 200 | 3.000000 |
| 171 | 201 | 4.000000 |
| 171 | 202 | 10.000000 |
| 172 | 130 | 15.000000 |
| 172 | 131 | 21.000000 |
| 172 | 132 | 19.000000 |
| 172 | 176 | 10.000000 |
| 172 | 183 | 5.000000 |
| 172 | 202 | 10.000000 |
| 173 | 130 | 15.000000 |
| 173 | 131 | 21.000000 |
| 173 | 132 | 19.000000 |
| 174 | 132 | 19.000000 |
| 175 | 182 | 5.000000 |
| 175 | 183 | 5.000000 |
| 175 | 202 | 10.000000 |
| 175 | 203 | 7.000000 |
| 176 | 172 | 29.000000 |
| 176 | 202 | 10.000000 |
| 176 | 203 | 7.000000 |
| 183 | 172 | 29.000000 |
| 183 | 175 | 11.000000 |
| 183 | 187 | 1.000000 |
| 202 | 171 | 14.000000 |
| 202 | 172 | 29.000000 |
| 202 | 175 | 11.000000 |
| 202 | 176 | 10.000000 |
| 202 | 177 | 7.000000 |
| 202 | 182 | 5.000000 |
| 202 | 223 | 1.000000 |
| Step 4 | | |
| 128 | 101 | 4.000000 |
| 128 | 102 | 4.000000 |
| 128 | 103 | 15.000000 |
| 128 | 168 | 16.000000 |
| 128 | 169 | 42.000000 |
| 128 | 170 | 86.000000 |
| 129 | 103 | 15.000000 |
| 129 | 169 | 42.000000 |
| 129 | 170 | 86.000000 |
| 129 | 171 | 129.000000 |
| 130 | 140 | 155.000000 |
| 130 | 141 | 126.000000 |
| 130 | 145 | 38.000000 |
| 130 | 170 | 86.000000 |
| 130 | 171 | 129.000000 |
| 130 | 172 | 154.000000 |
| 130 | 173 | 135.000000 |
| 131 | 138 | 124.000000 |
| 131 | 139 | 183.000000 |
| 131 | 140 | 155.000000 |
| 131 | 141 | 126.000000 |
| 131 | 171 | 129.000000 |
| 131 | 172 | 154.000000 |
| 131 | 173 | 135.000000 |
| 132 | 138 | 124.000000 |
| 132 | 139 | 183.000000 |
| 132 | 140 | 155.000000 |
| 132 | 172 | 154.000000 |
| 132 | 173 | 135.000000 |
| 132 | 174 | 95.000000 |
| 133 | 137 | 94.000000 |
| 133 | 138 | 124.000000 |
| 133 | 139 | 183.000000 |
| 134 | 138 | 124.000000 |
| 134 | 139 | 183.000000 |
| 137 | 133 | 107.000000 |
| 138 | 131 | 213.000000 |
| 138 | 132 | 204.000000 |
| 138 | 133 | 107.000000 |
| 138 | 134 | 107.000000 |
| 138 | 152 | 137.000000 |
| 138 | 153 | 132.000000 |
| 139 | 131 | 213.000000 |
| 139 | 132 | 204.000000 |
| 139 | 133 | 107.000000 |
| 139 | 134 | 107.000000 |
| 139 | 150 | 98.000000 |
| 139 | 151 | 107.000000 |
| 139 | 152 | 137.000000 |
| 139 | 153 | 132.000000 |
| 140 | 130 | 160.000000 |
| 140 | 131 | 213.000000 |
| 140 | 132 | 204.000000 |
| 140 | 148 | 69.000000 |
| 140 | 149 | 79.000000 |
| 140 | 150 | 98.000000 |
| 140 | 151 | 107.000000 |
| 140 | 152 | 137.000000 |
| 140 | 153 | 132.000000 |
| 141 | 130 | 160.000000 |
| 141 | 131 | 213.000000 |
| 141 | 145 | 38.000000 |
| 141 | 146 | 26.000000 |
| 141 | 147 | 38.000000 |
| 141 | 148 | 69.000000 |
| 141 | 149 | 79.000000 |
| 142 | 104 | 4.000000 |
| 142 | 146 | 26.000000 |
| 142 | 147 | 38.000000 |
| 142 | 148 | 69.000000 |
| 142 | 149 | 79.000000 |
| 145 | 130 | 160.000000 |
| 145 | 141 | 126.000000 |
| 146 | 141 | 126.000000 |
| 146 | 142 | 38.000000 |
| 147 | 141 | 126.000000 |
| 147 | 142 | 38.000000 |
| 147 | 143 | 15.000000 |
| 148 | 140 | 155.000000 |
| 148 | 141 | 126.000000 |
| 148 | 142 | 38.000000 |
| 149 | 140 | 155.000000 |
| 149 | 141 | 126.000000 |
| 149 | 142 | 38.000000 |
| 150 | 139 | 183.000000 |
| 150 | 140 | 155.000000 |
| 151 | 139 | 183.000000 |
| 151 | 140 | 155.000000 |
| 151 | 156 | 47.000000 |
| 152 | 138 | 124.000000 |
| 152 | 139 | 183.000000 |
| 152 | 140 | 155.000000 |
| 152 | 156 | 47.000000 |
| 153 | 138 | 124.000000 |
| 153 | 139 | 183.000000 |
| 153 | 140 | 155.000000 |
| 153 | 157 | 29.000000 |
| 154 | 158 | 23.000000 |
| 155 | 159 | 17.000000 |
| 156 | 151 | 107.000000 |
| 156 | 152 | 137.000000 |
| 156 | 160 | 11.000000 |
| 157 | 153 | 132.000000 |
| 157 | 161 | 3.000000 |
| 157 | 194 | 3.000000 |
| 168 | 126 | 8.000000 |
| 168 | 127 | 19.000000 |
| 168 | 128 | 37.000000 |
| 168 | 197 | 3.000000 |
| 168 | 198 | 15.000000 |
| 168 | 199 | 19.000000 |
| 169 | 126 | 8.000000 |
| 169 | 127 | 19.000000 |
| 169 | 128 | 37.000000 |
| 169 | 129 | 64.000000 |
| 169 | 198 | 15.000000 |
| 169 | 199 | 19.000000 |
| 169 | 200 | 30.000000 |
| 170 | 128 | 37.000000 |
| 170 | 129 | 64.000000 |
| 170 | 130 | 160.000000 |
| 170 | 198 | 15.000000 |
| 170 | 199 | 19.000000 |
| 170 | 200 | 30.000000 |
| 170 | 201 | 36.000000 |
| 171 | 129 | 64.000000 |
| 171 | 130 | 160.000000 |
| 171 | 131 | 213.000000 |
| 171 | 200 | 30.000000 |
| 171 | 201 | 36.000000 |
| 171 | 202 | 89.000000 |
| 172 | 130 | 160.000000 |
| 172 | 131 | 213.000000 |
| 172 | 132 | 204.000000 |
| 172 | 176 | 112.000000 |
| 172 | 183 | 50.000000 |
| 172 | 202 | 89.000000 |
| 173 | 130 | 160.000000 |
| 173 | 131 | 213.000000 |
| 173 | 132 | 204.000000 |
| 174 | 132 | 204.000000 |
| 175 | 182 | 36.000000 |
| 175 | 183 | 50.000000 |
| 175 | 202 | 89.000000 |
| 175 | 203 | 51.000000 |
| 176 | 172 | 154.000000 |
| 176 | 202 | 89.000000 |
| 176 | 203 | 51.000000 |
| 177 | 202 | 89.000000 |
| 177 | 203 | 51.000000 |
| 178 | 203 | 51.000000 |
| 181 | 203 | 51.000000 |
| 181 | 204 | 24.000000 |
| 181 | 205 | 9.000000 |
| 181 | 209 | 7.000000 |
| 182 | 175 | 121.000000 |
| 182 | 202 | 89.000000 |
| 182 | 203 | 51.000000 |
| 182 | 204 | 24.000000 |
| 182 | 209 | 7.000000 |
| 182 | 213 | 9.000000 |
| 183 | 172 | 154.000000 |
| 183 | 175 | 121.000000 |
| 183 | 187 | 8.000000 |
| 184 | 188 | 4.000000 |
| 184 | 213 | 9.000000 |
| 184 | 216 | 2.000000 |
| 185 | 189 | 3.000000 |
| 186 | 190 | 2.000000 |
| 187 | 183 | 50.000000 |
| 187 | 191 | 1.000000 |
| 187 | 216 | 2.000000 |
| 187 | 217 | 1.000000 |
| 198 | 167 | 5.000000 |
| 198 | 168 | 16.000000 |
| 198 | 169 | 42.000000 |
| 198 | 170 | 86.000000 |
| 198 | 218 | 5.000000 |
| 198 | 219 | 2.000000 |
| 198 | 220 | 5.000000 |
| 199 | 168 | 16.000000 |
| 199 | 169 | 42.000000 |
| 199 | 170 | 86.000000 |
| 199 | 218 | 5.000000 |
| 199 | 219 | 2.000000 |
| 199 | 220 | 5.000000 |
| 199 | 221 | 9.000000 |
| 199 | 222 | 9.000000 |
| 200 | 169 | 42.000000 |
| 200 | 170 | 86.000000 |
| 200 | 171 | 129.000000 |
| 200 | 213 | 9.000000 |
| 200 | 218 | 5.000000 |
| 200 | 220 | 5.000000 |
| 200 | 221 | 9.000000 |
| 200 | 222 | 9.000000 |
| 201 | 170 | 86.000000 |
| 201 | 171 | 129.000000 |
| 201 | 221 | 9.000000 |
| 201 | 222 | 9.000000 |
| 201 | 223 | 15.000000 |
| 202 | 171 | 129.000000 |
| 202 | 172 | 154.000000 |
| 202 | 175 | 121.000000 |
| 202 | 176 | 112.000000 |
| 202 | 177 | 44.000000 |
| 202 | 182 | 36.000000 |
| 202 | 223 | 15.000000 |
| 203 | 175 | 121.000000 |
| 203 | 176 | 112.000000 |
| 203 | 177 | 44.000000 |
| 203 | 178 | 35.000000 |
| 203 | 179 | 22.000000 |
| 203 | 180 | 15.000000 |
| 203 | 181 | 18.000000 |
| 203 | 182 | 36.000000 |
| 204 | 179 | 22.000000 |
| 204 | 180 | 15.000000 |
| 204 | 181 | 18.000000 |
| 204 | 182 | 36.000000 |
| 204 | 209 | 7.000000 |
| 204 | 223 | 15.000000 |
| 223 | 7 | 1.000000 |
| 223 | 8 | 1.000000 |
| 223 | 9 | 1.000000 |
| 223 | 201 | 36.000000 |
| 223 | 202 | 89.000000 |
| 223 | 204 | 24.000000 |

**Step 5**

| | | |
|---|---|---|
| 7 | 44 | 2.000000 |
| 7 | 45 | 1.000000 |
| 7 | 46 | 2.000000 |
| 7 | 47 | 3.000000 |
| 7 | 220 | 90.000000 |
| 7 | 221 | 115.000000 |
| 7 | 222 | 117.000000 |
| 7 | 223 | 173.000000 |
| 8 | 44 | 2.000000 |
| 8 | 46 | 2.000000 |
| 8 | 47 | 3.000000 |
| 8 | 48 | 2.000000 |
| 8 | 222 | 117.000000 |
| 8 | 223 | 173.000000 |
| 8 | 224 | 19.000000 |
| 9 | 47 | 3.000000 |
| 9 | 48 | 2.000000 |
| 9 | 49 | 1.000000 |
| 9 | 222 | 117.000000 |
| 9 | 223 | 173.000000 |
| 9 | 224 | 19.000000 |
| ==101== | ==78== | ==4.000000== |
| 101 | 79 | 4.000000 |
| 101 | 80 | 8.000000 |
| 101 | 125 | 36.000000 |
| 101 | 126 | 127.000000 |
| 101 | 127 | 195.000000 |
| 101 | 128 | 418.000000 |
| 102 | 80 | 8.000000 |
| 102 | 83 | 4.000000 |
| 102 | 106 | 23.000000 |
| 102 | 126 | 127.000000 |
| 102 | 127 | 195.000000 |
| 102 | 128 | 418.000000 |
| 103 | 107 | 19.000000 |
| 103 | 127 | 195.000000 |
| 103 | 128 | 418.000000 |
| 103 | 129 | 701.000000 |
| 104 | 108 | 4.000000 |
| 104 | 142 | 406.000000 |
| 104 | 143 | 155.000000 |
| 126 | 99 | 12.000000 |
| 126 | 100 | 16.000000 |
| 126 | 101 | 68.000000 |
| 126 | 102 | 87.000000 |
| 126 | 165 | 46.000000 |
| 126 | 166 | 14.000000 |
| 126 | 167 | 104.000000 |
| 126 | 168 | 234.000000 |
| 126 | 169 | 428.000000 |
| 127 | 101 | 68.000000 |
| 127 | 102 | 87.000000 |
| 127 | 103 | 128.000000 |
| 127 | 165 | 46.000000 |
| 127 | 167 | 104.000000 |
| 127 | 168 | 234.000000 |
| 127 | 169 | 428.000000 |
| 128 | 101 | 68.000000 |
| 128 | 102 | 87.000000 |
| 128 | 103 | 128.000000 |
| 128 | 168 | 234.000000 |
| 128 | 169 | 428.000000 |
| 128 | 170 | 702.000000 |
| 129 | 103 | 128.000000 |
| 129 | 169 | 428.000000 |
| 129 | 170 | 702.000000 |
| 129 | 171 | 1009.000000 |
| 130 | 140 | 1508.000000 |
| 130 | 141 | 999.000000 |
| 130 | 145 | 414.000000 |
| 130 | 170 | 702.000000 |
| 130 | 171 | 1009.000000 |
| 130 | 172 | 1394.000000 |
| 130 | 173 | 1188.000000 |
| 131 | 138 | 1177.000000 |
| 131 | 139 | 1604.000000 |
| 131 | 140 | 1508.000000 |
| 131 | 141 | 999.000000 |
| 131 | 171 | 1009.000000 |
| 131 | 172 | 1394.000000 |
| 131 | 173 | 1188.000000 |
| 132 | 138 | 1177.000000 |
| 132 | 139 | 1604.000000 |
| 132 | 140 | 1508.000000 |
| 132 | 172 | 1394.000000 |
| 132 | 173 | 1188.000000 |
| 132 | 174 | 726.000000 |
| 133 | 137 | 599.000000 |
| 133 | 138 | 1177.000000 |
| 133 | 139 | 1604.000000 |
| 134 | 138 | 1177.000000 |
| 134 | 139 | 1604.000000 |
| 137 | 133 | 790.000000 |
| 138 | 131 | 1434.000000 |
| 138 | 132 | 1273.000000 |
| 138 | 133 | 790.000000 |
| 138 | 134 | 790.000000 |
| 138 | 152 | 920.000000 |
| 138 | 153 | 856.000000 |
| 139 | 131 | 1434.000000 |
| 139 | 132 | 1273.000000 |
| 139 | 133 | 790.000000 |
| 139 | 134 | 790.000000 |
| 139 | 150 | 730.000000 |
| 139 | 151 | 831.000000 |
| 139 | 152 | 920.000000 |
| 139 | 153 | 856.000000 |
| 140 | 130 | 1137.000000 |
| 140 | 131 | 1434.000000 |
| 140 | 132 | 1273.000000 |
| 140 | 148 | 534.000000 |
| 140 | 149 | 631.000000 |
| 140 | 150 | 730.000000 |
| 140 | 151 | 831.000000 |
| 140 | 152 | 920.000000 |
| 140 | 153 | 856.000000 |
| 141 | 130 | 1137.000000 |
| 141 | 131 | 1434.000000 |
| 141 | 145 | 414.000000 |
| 141 | 146 | 266.000000 |
| 141 | 147 | 391.000000 |
| 141 | 148 | 534.000000 |
| 141 | 149 | 631.000000 |
| 142 | 104 | 72.000000 |
| 142 | 146 | 266.000000 |
| 142 | 147 | 391.000000 |
| 142 | 148 | 534.000000 |
| 142 | 149 | 631.000000 |
| 143 | 104 | 72.000000 |
| 143 | 147 | 391.000000 |
| 145 | 130 | 1137.000000 |
| 145 | 141 | 999.000000 |
| 146 | 141 | 999.000000 |
| 146 | 142 | 406.000000 |
| 147 | 141 | 999.000000 |
| 147 | 142 | 406.000000 |
| 147 | 143 | 155.000000 |
| 148 | 140 | 1508.000000 |
| 148 | 141 | 999.000000 |
| 148 | 142 | 406.000000 |
| 149 | 140 | 1508.000000 |
| 149 | 141 | 999.000000 |
| 149 | 142 | 406.000000 |
| 150 | 139 | 1604.000000 |
| 150 | 140 | 1508.000000 |
| 151 | 139 | 1604.000000 |
| 151 | 140 | 1508.000000 |
| 151 | 156 | 530.000000 |
| 152 | 138 | 1177.000000 |
| 152 | 139 | 1604.000000 |
| 152 | 140 | 1508.000000 |
| 152 | 156 | 530.000000 |
| 153 | 138 | 1177.000000 |
| 153 | 139 | 1604.000000 |
| 153 | 140 | 1508.000000 |
| 153 | 157 | 310.000000 |
| 154 | 158 | 184.000000 |
| 155 | 159 | 150.000000 |
| 156 | 151 | 831.000000 |
| 156 | 152 | 920.000000 |
| 156 | 160 | 119.000000 |
| 157 | 153 | 856.000000 |
| 157 | 161 | 80.000000 |
| 157 | 194 | 56.000000 |
| 158 | 154 | 405.000000 |
| 158 | 162 | 54.000000 |
| 158 | 193 | 29.000000 |
| 158 | 194 | 56.000000 |
| 159 | 155 | 422.000000 |
| 159 | 163 | 31.000000 |
| 160 | 156 | 530.000000 |
| 160 | 164 | 14.000000 |
| 160 | 165 | 46.000000 |
| 161 | 157 | 310.000000 |
| 161 | 165 | 46.000000 |
| 167 | 125 | 36.000000 |
| 167 | 126 | 127.000000 |
| 167 | 127 | 195.000000 |
| 167 | 196 | 26.000000 |
| 167 | 197 | 58.000000 |
| 167 | 198 | 214.000000 |
| 168 | 126 | 127.000000 |
| 168 | 127 | 195.000000 |
| 168 | 128 | 418.000000 |
| 168 | 197 | 58.000000 |
| 168 | 198 | 214.000000 |
| 168 | 199 | 258.000000 |
| 169 | 126 | 127.000000 |
| 169 | 127 | 195.000000 |
| 169 | 128 | 418.000000 |
| 169 | 129 | 701.000000 |
| 169 | 198 | 214.000000 |
| 169 | 199 | 258.000000 |
| 169 | 200 | 364.000000 |
| 170 | 128 | 418.000000 |
| 170 | 129 | 701.000000 |
| 170 | 130 | 1137.000000 |
| 170 | 198 | 214.000000 |
| 170 | 199 | 258.000000 |
| 170 | 200 | 364.000000 |
| 170 | 201 | 386.000000 |
| 171 | 129 | 701.000000 |
| 171 | 130 | 1137.000000 |
| 171 | 131 | 1434.000000 |
| 171 | 200 | 364.000000 |
| 171 | 201 | 386.000000 |
| 171 | 202 | 722.000000 |
| 172 | 130 | 1137.000000 |
| 172 | 131 | 1434.000000 |
| 172 | 132 | 1273.000000 |
| 172 | 176 | 724.000000 |
| 172 | 183 | 366.000000 |
| 172 | 202 | 722.000000 |
| 173 | 130 | 1137.000000 |
| 173 | 131 | 1434.000000 |
| 173 | 132 | 1273.000000 |
| 174 | 132 | 1273.000000 |
| 175 | 182 | 382.000000 |
| 175 | 183 | 366.000000 |
| 175 | 202 | 722.000000 |
| 175 | 203 | 525.000000 |
| 176 | 172 | 1394.000000 |
| 176 | 202 | 722.000000 |
| 176 | 203 | 525.000000 |
| 177 | 202 | 722.000000 |
| 177 | 203 | 525.000000 |
| 178 | 203 | 525.000000 |
| 179 | 203 | 525.000000 |
| 179 | 204 | 263.000000 |
| 179 | 205 | 139.000000 |
| 180 | 203 | 525.000000 |
| 180 | 204 | 263.000000 |
| 180 | 205 | 139.000000 |
| 181 | 203 | 525.000000 |
| 181 | 204 | 263.000000 |
| 181 | 205 | 139.000000 |
| 181 | 209 | 97.000000 |
| 182 | 175 | 801.000000 |
| 182 | 202 | 722.000000 |
| 182 | 203 | 525.000000 |
| 182 | 204 | 263.000000 |
| 182 | 209 | 97.000000 |
| 182 | 213 | 94.000000 |
| 183 | 172 | 1394.000000 |
| 183 | 175 | 801.000000 |
| 183 | 187 | 92.000000 |
| 184 | 188 | 46.000000 |
| 184 | 213 | 94.000000 |
| 184 | 216 | 40.000000 |
| 185 | 189 | 31.000000 |
| 186 | 190 | 27.000000 |
| 187 | 183 | 366.000000 |
| 187 | 191 | 21.000000 |
| 187 | 216 | 40.000000 |
| 187 | 217 | 31.000000 |
| 188 | 184 | 125.000000 |
| 188 | 216 | 40.000000 |
| 188 | 217 | 31.000000 |
| 189 | 185 | 86.000000 |
| 190 | 186 | 88.000000 |
| 190 | 194 | 56.000000 |
| 191 | 187 | 92.000000 |
| 191 | 217 | 31.000000 |
| 194 | 157 | 310.000000 |
| 194 | 158 | 184.000000 |
| 194 | 190 | 27.000000 |
| 196 | 166 | 14.000000 |
| 196 | 167 | 104.000000 |
| 197 | 167 | 104.000000 |
| 197 | 168 | 234.000000 |
| 197 | 219 | 48.000000 |
| 198 | 167 | 104.000000 |
| 198 | 168 | 234.000000 |
| 198 | 169 | 428.000000 |
| 198 | 170 | 702.000000 |
| 198 | 218 | 83.000000 |
| 198 | 219 | 48.000000 |
| 198 | 220 | 90.000000 |
| 199 | 168 | 234.000000 |
| 199 | 169 | 428.000000 |
| 199 | 170 | 702.000000 |
| 199 | 218 | 83.000000 |
| 199 | 219 | 48.000000 |
| 199 | 220 | 90.000000 |
| 199 | 221 | 115.000000 |
| 199 | 222 | 117.000000 |
| 200 | 169 | 428.000000 |
| 200 | 170 | 702.000000 |
| 200 | 171 | 1009.000000 |
| 200 | 213 | 94.000000 |
| 200 | 218 | 83.000000 |
| 200 | 220 | 90.000000 |
| 200 | 221 | 115.000000 |
| 200 | 222 | 117.000000 |
| 201 | 170 | 702.000000 |
| 201 | 171 | 1009.000000 |
| 201 | 221 | 115.000000 |
| 201 | 222 | 117.000000 |
| 201 | 223 | 173.000000 |
| 202 | 171 | 1009.000000 |
| 202 | 172 | 1394.000000 |
| 202 | 175 | 801.000000 |
| 202 | 176 | 724.000000 |
| 202 | 177 | 430.000000 |
| 202 | 182 | 382.000000 |
| 202 | 223 | 173.000000 |
| 203 | 175 | 801.000000 |
| 203 | 176 | 724.000000 |
| 203 | 177 | 430.000000 |
| 203 | 178 | 365.000000 |
| 203 | 179 | 196.000000 |
| 203 | 180 | 159.000000 |
| 203 | 181 | 214.000000 |
| 203 | 182 | 382.000000 |
| 204 | 179 | 196.000000 |
| 204 | 180 | 159.000000 |
| 204 | 181 | 214.000000 |
| 204 | 182 | 382.000000 |
| 204 | 209 | 97.000000 |
| 204 | 223 | 173.000000 |
| 205 | 179 | 196.000000 |
| 205 | 180 | 159.000000 |
| 205 | 181 | 214.000000 |
| 205 | 209 | 97.000000 |
| 205 | 226 | 26.000000 |
| 209 | 181 | 214.000000 |
| 209 | 182 | 382.000000 |
| 209 | 204 | 263.000000 |
| 209 | 205 | 139.000000 |
| 209 | 213 | 94.000000 |
| 213 | 182 | 382.000000 |
| 213 | 184 | 125.000000 |
| 213 | 200 | 364.000000 |
| 213 | 209 | 97.000000 |
| 213 | 217 | 31.000000 |
| 213 | 218 | 83.000000 |
| 216 | 184 | 125.000000 |
| 216 | 187 | 92.000000 |
| 216 | 188 | 46.000000 |
| 216 | 212 | 18.000000 |
| 217 | 187 | 92.000000 |
| 217 | 188 | 46.000000 |
| 217 | 191 | 21.000000 |
| 217 | 213 | 94.000000 |
| 218 | 198 | 214.000000 |
| 218 | 199 | 258.000000 |
| 218 | 200 | 364.000000 |
| 218 | 213 | 94.000000 |
| 218 | 214 | 19.000000 |
| 219 | 4 | 7.000000 |
| 219 | 5 | 17.000000 |
| 219 | 197 | 58.000000 |
| 219 | 198 | 214.000000 |
| 219 | 199 | 258.000000 |
| 219 | 214 | 19.000000 |
| 220 | 4 | 7.000000 |
| 220 | 5 | 17.000000 |
| 220 | 6 | 24.000000 |
| 220 | 7 | 40.000000 |
| 220 | 198 | 214.000000 |
| 220 | 199 | 258.000000 |
| 220 | 200 | 364.000000 |
| 221 | 5 | 17.000000 |
| 221 | 6 | 24.000000 |
| 221 | 7 | 40.000000 |
| 221 | 199 | 258.000000 |
| 221 | 200 | 364.000000 |
| 221 | 201 | 386.000000 |
| 221 | 210 | 25.000000 |
| 222 | 6 | 24.000000 |
| 222 | 7 | 40.000000 |
| 222 | 8 | 27.000000 |
| 222 | 9 | 27.000000 |
| 222 | 199 | 258.000000 |
| 222 | 200 | 364.000000 |
| 222 | 201 | 386.000000 |
| 223 | 7 | 40.000000 |
| 223 | 8 | 27.000000 |
| 223 | 9 | 27.000000 |
| 223 | 201 | 386.000000 |
| 223 | 202 | 722.000000 |
| 223 | 204 | 263.000000 |
| 224 | 8 | 27.000000 |
| 224 | 9 | 27.000000 |
| 224 | 10 | 3.000000 |
| 224 | 228 | 3.000000 |
| 225 | 229 | 2.000000 |
| 226 | 205 | 139.000000 |
| 226 | 230 | 1.000000 |
| 226 | 231 | 1.000000 |
| 226 | 232 | 1.000000 |

**Step 6**

| | | |
|---|---|---|
| 76 | 98 | 5.000000 |
| 77 | 98 | 27.000000 |
| ==78== | ==98== | ==114.000000== |
| 93 | 98 | 12.000000 |
| 124 | 98 | 191.000000 |

**Shortest Path Residues:**

**Path 1**: 132, 173, 130, 128, 101, 78, 98
**Path 2**: 132, 172, 130, 128, 101, 78, 98
**Path 3**: 132, 140, 130, 128, 101, 78, 98

## 3. Residues of Shortest Path 1:

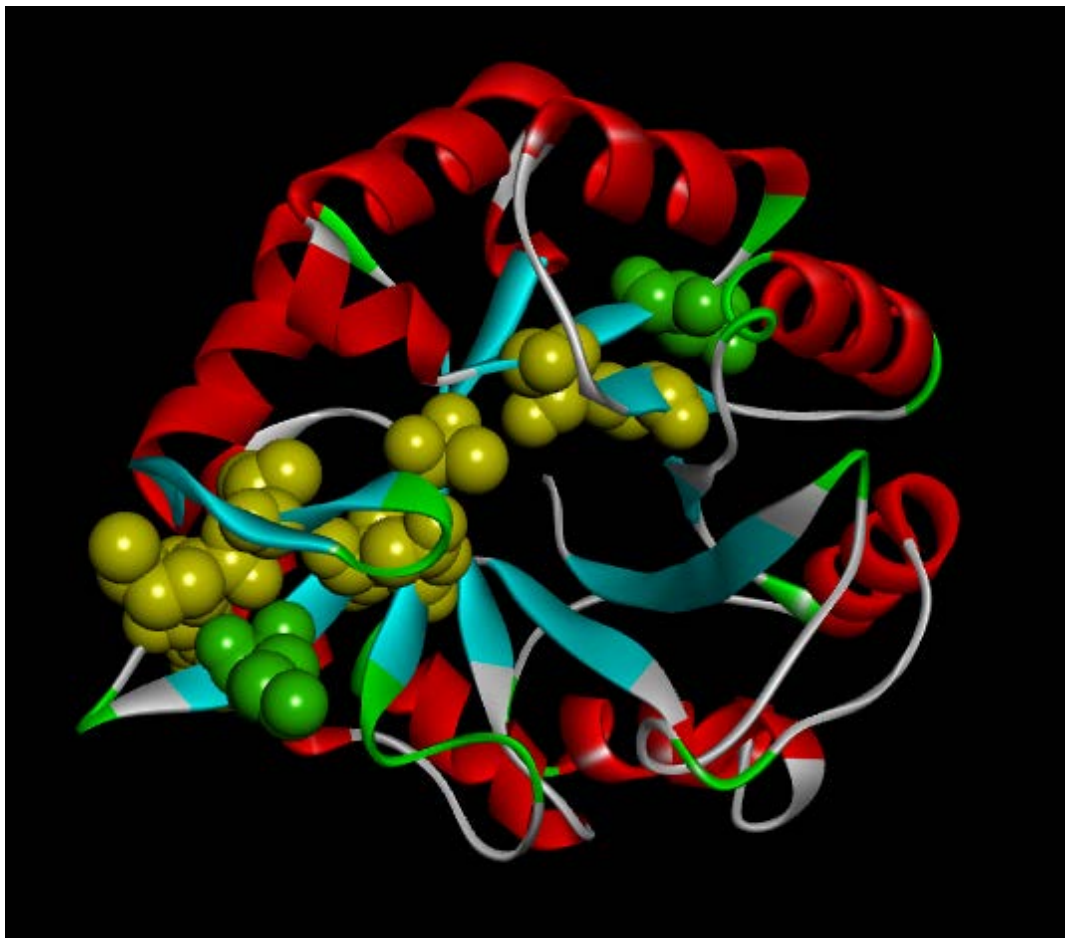

Figure S1. Shortest path residues highlighted in yellow. Terminal residues are in green.